\documentclass{mpe_report}

\usepackage{psfig,graphicx,epsfig}
\usepackage{color}
\usepackage{amsmath,amssymb,epic,eepic,array}

\begin{document}

\pagenumbering{arabic}
\setcounter{page}{68}

\renewcommand{\FirstPageOfPaper }{ 68}\renewcommand{\LastPageOfPaper }{ 71}

\title{Amazing properties of giant pulses and the nature of pulsar's radio emission}
\author{V.~Soglasnov\inst{}}
\institute{Astro Space Center of the Lebedev Physical Institute,
Profsoyuznaya str. 84/32, 117997 Moscow, Russia}
\maketitle

\begin{abstract}
 For comprehensive studying of giant pulses (GPs) from the Crab pulsar
and the original millisecond pulsar (MSP) B1937+21 (J1939+2134),
we conducted multifrequency observations over the last few years. They show that giant pulses may
be improbably bright, $10^5, 10^6$~Jy and more, they have extra
ordinal spectra and polarization. EM energy concentrated in such
strong pulse is high enough to accelerate particles up to Lorenz
factor $\gamma~\sim 10^4 - 10^6$, since giant pulses may play an important
role in physics of pulsar's magnetosphere.
\end{abstract}

\section{Introduction}

Giant pulses (GPs) are a very specific class of single radio pulses
observed in several pulsars which have power law intensity
distribution, while normally the distribution is of exponential
type. Therefore, sometimes one can detect a pulse exceeding by many
times a normal pulse intensity. However, the time of occurrence of
such events is unpredictable, it makes very difficult their search
and investigation. At the moment giant pulses are detected or
suspected in nearly a dozen of pulsars, but only two, the Crab
pulsar and the original millisecond pulsar (MSP) B1937+21
(J1939+2134), have the rate of GP occurrence sufficiently high for
good statistics and more or less detail study of this phenomenon.
During few last years we conducted many observations of giant pulses
from these pulsars over a wide frequency range from 20~MHz to 5~GHz,
single and multi frequency/station. The observations were made in
collaboration with many observatories and observers: Kalyazin
observatory with 64-m dish, Yu.~Ilyasov V.~Oreshko; Arecibo 305-m
telescope, T.~Hankins; 100-m GBT, Yu.~Kovalev, F~Ghigo (NRAO GB);
ARO (40-m dish in Canada), N.~Bartel, W.~Cannon, A.~Novikov (York
University); WSRT, B.~Stappers (NFRA); UTR-2 decametric telescope,
O.~Ulyanov, V.~Zakharenko (Ukrainian Institute of Radio Astronomy). An
important feature of these observations is a long time continuous
record with high time resolution (the latter is necessary because
giant pulses are very short, see below, paragraph~\ref{intens}). We
used standard VLBI terminals Mk5A (Kovalev et~al. 2005), Canadian S2 and Japanese K5 as
high performance and large capacity data acquisition systems. They
provide time resolution 8--16~ns and 6--12 hours continuous record.
Raw data were then encoded and processed for coherent dedispersion.

In section~\ref{obs} we describe briefly main results of these
observations, the most important is that the peak flux density of
giant pulse may reach improbable huge value $10^5-10^6$~Jy and more.
The interaction of such strong EM wave with plasma particles is very
specific (section~\ref{em}). In particular, a strong wave may work
as effective particles accelerator. In the
sections~\ref{wh},~\ref{aft} we discuss a probable origin of giant
pulses and their role in physical processes inside magnetosphere.

\section{Observational properties}
\label{obs}
    \subsection{Waveform and time duration}
        \label{Wform}

\begin{figure}
\centerline{\psfig{file=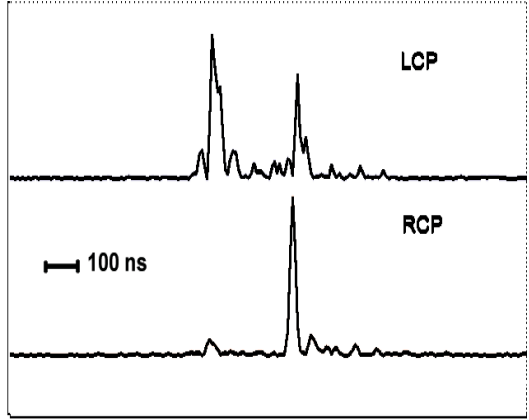,width=7.0cm,clip=} }
\caption{Example of rare occurred multi component giant pulse from
MSP B1937+21, f=2.2~GHz, as it seen in two polarization channels.
The first component has nearly total circular polarization of one
sign (LCP). A leading edge of the second component also 100\%
circularly polarized, but of opposite sign (RCP). Polarization
changes the sign at the middle of second component. \label{image}}
\label{msp}
\end{figure}

 Giant pulses from MSP B1937+21 initially are extremely short, less
than 10 ns~(\cite{kondrat2006,soglasnov2004}).
Their apparent waveform and time duration are caused completely by
interstellar scattering. Only several pulses from the analyzed few
thousands can be suspected as having some structure other than
scattering waveform (Fig.~\ref{msp}).

\begin{figure}
\centerline{\psfig{file=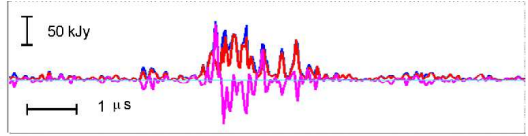,width=7.0cm,clip=} }
\caption{Example of extent giant pulse from the Crab pulsar,
detected at 2.2 GHz, $\Delta\nu=16 \rm{MHz}$, displayed in total
intensity (blue), linear (red) and circular(magenta) polarization.
\label{image}}
\label{long}
\end{figure}

\begin{figure}
\centerline{\psfig{file=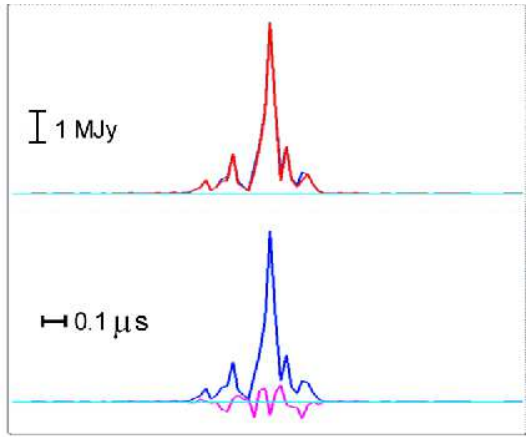,width=7.0cm,clip=} }
\caption{Example of extremely bright linearly polarized giant pulse
from the Crab pulsar,$S_{peak}=5.4 \rm{MJy}$, detected at 2.2 GHz,
$\Delta\nu=16 \rm{MHz}$, displayed in total intensity (blue), linear
(red) and circular(magenta) polarization. \label{image}}
\label{strong1}
\end{figure}

\begin{figure}
\centerline{\psfig{file=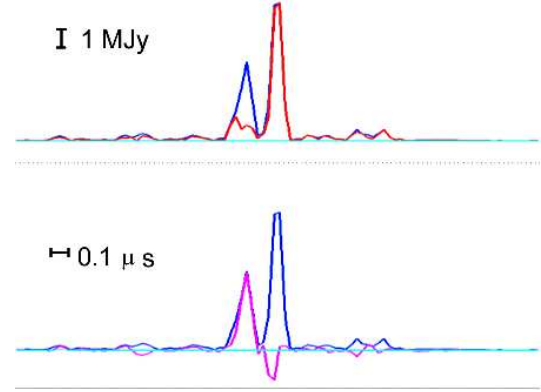,width=7.0cm,clip=} }
\caption{Example of extremely bright double giant pulse from the Crab
pulsar, $S_{peak}=4.2~\rm and~7.3~MJy$, detected at 2.2 GHz,
displayed in total
intensity (blue), linear (red) and circular(magenta) polarization.
Both components have almost 100\% polarization, the first circular
and the second linear. \label{image}}
\label{strong2}
\end{figure}

Waveform and duration of the Crab giant pulses depend on their
strength. Weak and medium GPs generally have complex structure. The
total duration may reach few decades of microseconds
(Fig~\ref{long}). However, giant pulses with peak flux density
exceeding some critical value (300~kJy at 1.4 GHz) are also very
short. They consist of one or two narrow peaks, sometimes with
weaker pedestal of $1-2~\mu s$ duration
(Fig.~\ref{strong1},~\ref{strong2}).

Giant pulses from MSP and those strong from the Crab are very similar, in spite of a
great difference of magnetic field strength and other physical
conditions. In both pulsars strong events intrinsically are
extremely short, we could not resolve them. Perhaps, weak extent
giant pulses from MSP also exist but they are below detection
threshold, because the pulsar B1937+21 is more distant and initially weaker.

\subsection{Intensity}
    \label{intens}

 Obviously power law distribution of giant pulse intensity can not be
continued up to zero as well as to the infinity, it {\it must} have
cutoff both at low and high energies (or, at least, strong break in
power index: flattening at low and steepening at high energies). Low
intensity fattening of the distribution for Crab giant pulses was
recently detected by ~\cite{pop2006}: $E_{break}=1000 - 3500~
\rm{Jy\cdot \mu s}$ at 1.2 GHz \footnote{In case of extent pulses,
flux density integrated over the whole pulse duration, rather it's
peak value, is a measure of pulse strength. Observers name it
``energy" and use the units of strange but convenient in practice
dimension [${\rm~Jy\cdot~\mu~s}$].}.

In case of MSP the limit is far below GP detection threshold, since
it cannot be observed directly. However, it may be derived under
adoption that power law distribution is valid up to this limit
~(\cite{soglasnov2004}): $S_{min}$=16~Jy (main pulse), $S_{min}$=5~Jy
(interpulse) at 1.2 and 1.6 GHz.

As for high intensity limit, the situation seems to be rather
intriguing. Now there are no indications on the deviation from power
law distribution up to the improbable high intensity. Giant pulses
from the Crab pulsar with peak flux density exceeding million jansky
(1 MJy) were detected at 2.2 and 1.4 GHz in observations conducted
in 2005 year (Fig.~\ref{strong1},~\ref{strong2}). Such events are
not something exclusive, they normally occurred in each session if
the observing time was sufficiently long, in accordance with power
law distribution. At the moment we have caught a half of hundred
``millionaires" over total 30 hours. It is difficult to provide much
more longer observing time which is needed to detect more intense
pulses, because of huge volume of data. We tested at Kalyazin the system for long
time multi frequency monitoring of the Crab total power without any
dispersion removal, which can provide detection the strongest giant pulses. Few days of
probe observations show tat the system works properly. We plan to
start regular monitoring at 5, 1.4 and 0.6~GHz in September 2006.

\subsection{Polarization}
\label{pol}

 The emission of narrow giant pulses is strongly polarized, up to
100~\% of linear, circular (may be of both signs), elliptical,
either pure or variously mixed. In itself it is not very
surprisingly, because their extremely short duration means that the
emission originates from very compact region, since
particles are emitting under the same physical conditions. What is really
marvelous, that giant pulses show rapid changes of polarization over
a very short time interval. Thus, two peaks separated by only few
decades of nanoseconds (since close spatially) may have quite
different polarization (Fig.~\ref{msp},~\ref{strong2}). Moreover,
polarization may change rapidly inside a single narrow component,
from circular to linear and opposite, or/and circular polarization may
change the sign. It seems improbable that the conditions change
considerably at so small scale ($\sim 1 - 10$~m). The only
explanation we can propose at the moment is that the emission may be
produced by stratified particles of opposite sign of charge
(e.g. electrons and positrons), as a result, few spikes alternatively RCP
and LCP polarized are generating, many of them are undistinguishable
with our time resolution. Similar picture was observed by
~\cite{hankins2003} at higher frequencies.

\subsection{Spectra}
\label{spectr}
 In average, giant pulses have nearly power law
spectra. For MSP the spectral index equals -2.8. For the Crab
pulsar, the spectrum becomes slightly flattened at higher
frequencies, it changes from $-2.7$ (at frequencies below 1GHz) to
$-1.8$ (between $1.4 - 2.2 \rm GHz$).

Instant spectra of single pulses, obtained in simultaneous multi
frequency sessions, are very different. The spectral index change
within wide limits from $+0.4$ to $-4.0$ ~\cite{popov2006a}. Only
one third of events occur simultaneously at widely spaced
frequencies. Partially it can be explained by interstellar
scintillations,
but it is not a main reason, as it follows from simultaneous
observations of the Crab at 600 and 111~MHz, where the receiver
bandpass exceed greatly decorrelation band, since interstellar
scintillations do not affect considerably on the apparent strength
of GPs. Detail analysis shows that spectra of single giant pulses
cover a wide frequency range but not continuously. They represent
chains of spots or bands, which parameters are inconsistent with
usual ISS; probably they are intrinsic gant pulse structure. The
most remarkable examples are quite exclusive spectra of the Crab
giant interpulses with discrete regular structure observed at
Arecibo at frequencies higher than 5GHz ~\cite{eilek2006}.

\section{Giant pulses and pulsar's physics}
\label{phys}
\subsection{Giant pulses as strong EM waves}
\label{em}
 Megajansky intensity of GPs detected in the Crab pulsar
looks rather impressive, however, are these pulses really giant with
physical point of view? There are several different criteria for the
strong EM wave. The top is Schwinger's quantum limit, when the field
strength of EM wave exceeds a critical value $4.4 \times
10^{13}\rm~Gs$. Such wave can create $e^+ e^-$ pairs. It is
convenient to measure the wave field strength $E_w$ in (angular)
frequency units $\omega_w\equiv~eE_w/m_ec=eH_w/m_ec$, then the
Schwinger's limit can be written as $\hbar\omega_w>>m_ec^2$. The
next criterium is condition $\omega_w>\omega$, where $\omega$ is
wave frequency. It means that the motion of charged particles under
the action of EM wave becomes relativistic, such wave accelerates
particles nearly up to the speed of light over a single wave cycle.

From the observed peak flux density $S_{Jy}$ we can estimate
$\omega_w$ at the distance $l_{cm}$ from the region of GP emission
as
$$\omega_w \approx 4.9\cdot 10^{12}{{L_{kpc}\over {l_{cm}}}\sqrt{S_{Jy}\Delta
\nu}},$$ where $L_{kpc}$ is the distance from the Earth to the
pulsar in kiloparsecs, $\Delta\nu$ is the frequency band. This
estimate is obtained without any arbitrary adoption such as
dimension of GP source, beam pattern of the emission {\it etc.}, it
based only on the inverse square law, which is valid at least up to
the boundary of wave zone. Even strongest ``MegaJansky" Crab pulses
detected at 2.3 and 1.4 GHz are far below the Schwinger's limit,
however, they satisfy to the condition $\omega_w>\omega$, which is
valid up to the distance $10^{10} - 10^{11}\rm{cm}$ from the
emitter, or $\simeq100$ radii of the light cylinder. Since inside
magnetosphere pulses are ``really giant": if suppose that they are
emitted near the star's surface, the ratio $\omega_w/\omega\geq100$
even at light cylinder, $10^3 - 10^6$ near the emitter (at
$10^4 - 10^6\rm{cm}$). This ratio may be much more for giant pulses at low
frequency 23 MHz: if adopt the true (non-scattered) pulse duration
of order $\sim 1\mu s$,~~$\omega_w/\omega\sim10^4$ at light cylinder
and$\sim10^6 - 10^7$ near the emitter.

Under condition $\omega_w/\omega\>>1$ the interaction between EM
wave and plasma particles becomes very specific. The wave
accelerates particles up to the Lorenz factor
$\gamma\sim\omega_w/\omega$. Particles emit secondary waves at {\it
different} frequencies and in {\it different} directions than the
incident wave. Thus, in the simplest case of circularly polarized
wave, trajectory of particles is a circle of radius $\lambda/2\pi$
($\lambda$ is wavelength), the plane of the circle coincides with
the wavefront. Particles emit at frequency
$\omega_{em}\sim\omega\gamma^2\sim\omega_w^2/\omega$ under right
angle to the direction of the incident wave propagation within a
small angle $\sim\gamma^{-1}\sim\omega/\omega_w$.

The case of linearly polarized wave is more complicate. Particles
move along trajectory of eight-like shape in the plane which is
normal to the wave front, the trajectory crosses itself at the
center under angle $51^{\circ}.1$,$\gamma$-factor changes along
trajectory from 1.03$\omega_w/\omega$ in the middle to
0.36$\omega_w/\omega$ at the edges of trajectory.

Some evidence of the existence of such ``perpendicular" emission is a
dramatic transformation of the Crab pulsar profile at high
frequencies~\cite{moffet1996}, where two strong wide components HFC1 and HFC2 appear,
they are spaced by $51^{\circ} - 54^{\circ}$ in longitude, the
longitude of HFC2 is $-90^{\circ}$ relative the main pulse. They may
be interpreted as counterparts of giant pulses emitted at lower
frequencies, as the sum of large number of short duration bursts
(``secondary giant pulses"), in accordance with the results obtained
by ~\cite{aga2006}. Final confirmation may be obtained from
polarization measurement of single pulse HFC emission with high time
resolution.

In presence the magnetic field of strength {\it H}, in case of wave
propagation along field lines (the case important for pulsars),
the particles accelerated by the wave are orbiting with frequency
${\omega_H\omega_w/\omega}$, where $\omega_H\equiv{{eH}/{m_e
c}}=1.78 \times 10^7~H$. They emit synchrotron radiation at frequency
$\omega_{em}\sim\omega_H\omega_w^2/\omega^2$. Near star's surface
$H\sim10^{12}\rm{Gs}$ (Crab) and $\sim10^8 \rm{Gs}$ (MSP B1937+21),
since synchrotron quants have energy $\sim 10^{12}$ and $\sim
10^{16} \rm{eV}$ correspondingly, which is far above threshold
$e^+e^-$ pair creation by quants in magnetic fields near surface of
these pulsars. This mechanism is much more effective than
traditional curvature radiation, because synchrotron quants are
emitting under right angle to the field lines, while curvature
quants are emitting along the tangent field line. Thus, giant pulse
may work as effective particles' accelerator and is able to induce
cascade pair creation. Moreover, under some conditions giant pulse
becomes self-generating, see below, section~\ref{wh}.

\subsection{Where and how giant pulses are emitting}
\label{wh}

 All stated above are observational results and their
direct consequences. Now we concern briefly more speculative
question on the origin of giant pulses.

 In the frame of standard model of pulsar's magnetosphere
it seems rather unlikely that giant pulses originate from the region
located far from the star's surface, where the energy of radiation
per unit volume, corresponding to megajansky peak flux, exceeds by
many orders the density of plasma energy. By this reason giant
pulses surely cannot arise near the light cylinder. The only place
where sufficient number of particles have sufficient energy to
produce giant pulse is a polar cap near star's surface. Perhaps, a
cause of giant pulses is the gap discharge, they are directly
observed effect of the discharge.

If so, we can suppose the follow scenario. The first giant pule may
be born when the gap potential reaches some critical value, strong
currents begin to run across the polar cap generating strong EM
emission. As described above (in the section~\ref{em}), interaction
between strong EM wave (``giant pulse") and charged particles leads
in a strong magnetic field to the $e^+e^-$ pair creation. Because
this mechanism is very effective, immediately cascade pair creation
develops very fast in a small volume, begetting rapidly rising
volume charge, in turn, it generates strong EM emission (``secondary"
giant pulse), which accelerates particles, particles emit high
energy synchrotron quants which produce $e^+e^-$ pairs, and so on.

\subsection{What will happen afterwards}
\label{aft}

 The rate of pair production drops rapidly when the wave and
particles move away from the star's surface both because of the
emission dilution and decreasing the magnetic field strength.
However, as before, the wave is capable to accelerate particles over
the whole path inside magnetosphere up to light cylinder. It leads
to the development of strong plasma instabilities, as a result, a
``normal" pulsed radioemission is generating by usual plasma
mechanism.

Pulsars with long period have a large light cylinder radius, the
energy of EM wave (``giant pulse") propagating over such long path,
converts entirely to the "normal" pulsed emission. It is the reason
why giant pulses are not observed in the majority of long period
pulsars.


\begin{acknowledgements}
The author thanks all participants of observations. Especial thanks
Tim Hankins, Joanna Rankin, Norbert Bartel, Yurii Koalev for
exclusively useful discussions and financial support for visiting
observatories and providing 1-TB Big LaCie Disks for storing the
observational data. This work is supported by the Russian Foundation
for Basic Research (project number 04-02-16384) and the Presidium of
the Russian Academy of Sciences, the project ``Origin and evolution
of stars and galaxies''. We gratefully acknowledge the support by
the WE-Heraeus foundation.
\end{acknowledgements}

      \clearpage


\begin{thebibliography}{9}

\bibitem[Hankins et~al. (2003)]
{hankins2003}
Hankins, T.~H., Kern, J.~S., Weatherall, J.~C., \& Eilek, J.~A.
2003, Nature, 422, 141

\bibitem[(Eilek \& Hankins 2006)]{eilek2006}
Eilek J.~A. \& Hankins T.~H., 2006, These proceedings, "Radio
Emission Physics in the Crab Pulsar"

\bibitem[Kondratiev et~al. 2006]{kondrat2006}
V.I.~Kondratiev, M.V.~Popov, V.A.~Soglasnov, Y.Y.~Kovalev,
N.~Bartel, \and F.~Ghigo, 2006, These proceedings, "Detailed study
of giant pulses from the millisecond pulsar B1937+21"

\bibitem[Sovakovska et~al. (2006)]{aga2006}
A.~Slovakovska, A.~Jessner, G.~Kanbach, \and B.~Klein, 2006, These
proceedings, "Comparison of giant pulses in young and millisecond
pulsars"

\bibitem[Popov \& Stappers (2006)]{pop2006}
M.~Popov \& B.~Stappers 2006, A\&A, in preparation

\bibitem[(Kovalev et~al. 2005)]
{kovalev2005}
Kovalev, Y.~Y., Ghigo, F., Kondratiev, V.~I., et al.\ 2005,
"Observing at the GBT with VLBA$+$Mark5A and VLBA$+$S2 backends", GBT
Commissioning Memo~236; available at 
{\tt http://wiki.gb.nrao.edu/pub/Knowledge/GBTMemos/ GBT\_Mark5A\_S2.pdf}

\bibitem[(Popov et~al. 2006b)]
{popov2006a}
Popov, M.~V., Soglasnov, V.~A., Kondratiev, V.~I., et al.
2006b, ARep, 50, 55, transl. from: AZh, 2006, 83, 62

\bibitem[(Moffet \& Hankins 1996)]{moffet1996}
Moffet,~D.~A., \& Hankins,~T.~H. 1996, ApJ 468, 779


\bibitem[Soglasnov et~al. 2004]
{soglasnov2004}
Soglasnov, V.~A., Popov, M.~V., Bartel, et al.  2004, ApJ, 616, 439

\end{thebibliography}
\end{document}